\documentclass[slac_one]{revtex4}
\usepackage{epsfig}
\usepackage{graphicx}
\usepackage{fancyhdr}
\def\DESepsf(#1 width #2){\epsfxsize=#2 \epsfbox{#1}}

\newcommand{\be}{\begin{eqnarray}}
\newcommand{\en}{\end{eqnarray}}
\newcommand{\ov}{\overline}
\newcommand{\A}{{\cal A}}
\newcommand{\B}{{\cal B}}
\newcommand{\Sc}{{\cal S}}

\newcommand{\U}{{\cal U}}
\newcommand{\V}{{\cal V}}
\begin{document}

\title{Rescattering effects in charmless $\overline B_{u,d,s}\to P P$ decays}

%

\author{Chun-Khiang Chua}
\affiliation{Department of Physics, Chung Yuan Christian
University, Chung-Li, Taiwan 32023, Republic of China}

\begin{abstract}
Final-state interaction effects in charmless $\ov B_{u,d,s}\to PP$
decays are studied. We investigate the $\ov B{}^0\to
\pi^+\pi^-,\pi^0\pi^0$ rates and the $K\pi$ direct CP violations,
which lead to the so-called $K\pi$ puzzle in CP violation. Our
main results are as follows: (i) Results are in agreement with
data in the presence of FSI. (ii) For $\ov B$ decays, the
$\pi^+\pi^-$ and $\pi^0\pi^0$ rates are suppressed and enhanced
respectively by FSI. (iii) The FSI has a large impact on direct CP
asymmetries ($\A$) of many modes. (iv) The deviation($\Delta \A$)
between $\A(\ov B{}^0\to K^-\pi^+)$ and $\A(B^-\to K^-\pi^0)$ can
be understood in the FSI approach. (v) Sizable and complex
color-suppressed tree amplitudes, which are crucial for the large
$\pi^0\pi^0$ rate and $\Delta\A$, are generated through exchange
rescattering. The correlation of the ratio
$\B(\pi^0\pi^0)/\B(\pi^+\pi^-)$ and $\Delta \A$ is studied.
(vi)~The $B^-\to \pi^-\pi^0$ direct CP violation is very small and
is not affected by FSI. (vii)~Several $\ov B_s$ decay rates are
enhanced. In particular, the $\eta'\eta'$ branching ratio is
enhanced to the level of $1.0\times 10^{-4}$. 
(viii)~Time-dependent CP asymmetries $S$ in $\ov B_{d,s}$ decays
are studied. The $\Delta S(\ov B {}^0\to K_S\eta')$ is very small
($\leq 1\%$). We found that the asymmetries $|S|$ for $\ov B
{}^0_s\to \eta\eta$, $\eta\eta'$ and $\eta'\eta'$ decays are all
below $0.06$. CP asymmetries in these modes will be useful to test
the SM.
\end{abstract}

\maketitle

\thispagestyle{fancy}


\section{Introduction}

The study of $B$ decays provides many useful information of the
flavor sector of the Standard Model (SM)~\cite{data} and it also
enables us to search for possible New Physics effects. Recently
there are many interesting results~\cite{data}: (a) The
measurements of the time-dependent CP asymmetries in kaon and
charmonium final states give a rather precise value of
$\sin2\beta=0.681\pm0.025$~\cite{data}, where
$\beta/\phi_1=\arg(V^*_{td})$ with $V$ the
Cabbibo-Kobayashi-Mashikawa (CKM) matrix. In the SM,
time-dependent CP asymmetries in penguin dominated modes are
expected to be close to the $\sin2\beta$ value. Consequently,
these asymmetries are promising places to search for new physics
effects~\cite{review}. (b) Although $\A(B^-\to
K^-\pi^0)\simeq\A(\ov B{}^0\to K^-\pi^+)$ was expected in many
early theoretical
predictions~\cite{pQCD,Beneke:2001ev}, 
the recent measurements show $\A(K^-\pi^+)=(-9.8^{+1.2}_{-1.1})\%$
and $\A(K^-\pi^0)=(5.0\pm2.5)\%$~\cite{data}, giving 
$\A(K^-\pi^0)-\A(K^-\pi^+)=(14.8^{+2.7}_{-2.8})\%$, which is more
than $5\,\sigma$ from zero. This is the so-called $K\pi$ puzzle in
direct CP violation. (d) The observed large $\ov
B{}^0\to\pi^0\pi^0$ rate~\cite{data} is puzzling and posing
tension in many theoretical calculations. (e) Data for $\overline
B_s$ decays are starting to emerge from the Tevetron~\cite{data}
and from $B$ factories, and we anticipate
more to come in the near future. 
Measurements of rates and CP asymmetries in $\ov B_s$
decays will be useful in testing the SM and in searching for new
(physics) phases. In fact, recently, a claim on the evidence of
new physics effect in the $\ov B{}_s$ mixing was put
forward~\cite{UTfit}.  

\begin{figure}[t!]
 \centerline{\DESepsf(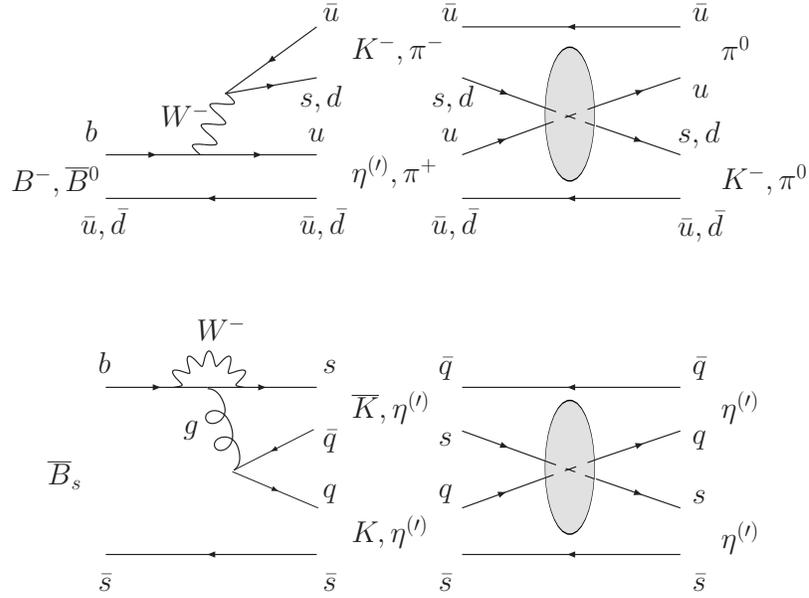 width 10cm)}
\caption{Exchange rescattering in $\ov B{}^0\to\pi^0\pi^0$,
$B^-\to K^-\pi^0$ and $\ov B_s\to\eta^{(\prime)}\eta^{(\prime)}$
decays.} \label{fig:re}
\end{figure}

Before we can claim on New Physics effects in the above puzzles,
it is important to investigate these processes within the SM
carefully. As shown in Fig.~\ref{fig:re}, final state interaction
(FSI) may affect $\ov B\to\pi^0\pi^0$ and $K^-\pi^0$ amplitudes
and, hence, may help to resolve the the $\pi\pi$ and $K\pi$
puzzles at the same time~\cite{Chua:2007cm}.
In~\cite{Chua:2007cm} we consider a FSI approach with both short-
and long-distance contributions, where the former are from
in-elastic channels and are contained in factorization amplitudes,
while the latter are from the residual rescattering among $PP$
states. We have
 \be
 A^{FSI}_i=\sum_{j=1}^n(\Sc_{res}^{1/2})_{ij} A^{fac}_j,
 \label{eq:master1}
 \en
where $i,j=1,\dots,n$ denote all charmless $PP$ states,
$A^{fac}_j$ is the factorization amplitude. The residual
rescattering effect is encoded in the $\Sc_{res}$ matrix. In
factorization approaches, the above $\Sc_{res}$ is taken to be
unity.
For the factorization amplitudes, we use those obtained in the QCD
factorization approach~\cite{Beneke:2003zv}.
Since strong interaction has (an approximate) SU(3) symmetry,
which is expected to be a good one at the $m_B$ rescattering scale
and, hence, can be used to constrain the form of $S_{res}$.
From SU(3) symmetry and the Bose-Einstein statistics, we have
 \be
 (\Sc_{res})^{1/2}
 =\sum_{a=1}^{27}|{\bf 27};a\rangle e^{i\delta_{27}}\langle {\bf 27};a|
  +\sum_{b=1}^{8}\sum_{p,q=8,8'}|p;b\rangle \U^{1/2}_{pq} \langle q;b|
  +\sum_{p,q=1,1'}|p;1\rangle \V^{1/2}_{pq} \langle q;1|,
 \label{eq:SU3decomposition}
 \en
where $\U$ and $\V$ are mixing matrices.
Experimental data is used to determine phases and mixing angles in
$\Sc_{res}$.
Note that similar formulas for $\ov B\to DP, \ov D P, PP$
rescatterings from SU(3) symmetry have been used in
\cite{quasielastic,Smith}.

\section{Results}

\begin{table}[t!]
\caption{ \label{tab:table-br} \label{tab:table-acp}
\label{tab:table-Bs} Branching ratios $\B$ 
and direct CP asymmetries $\A$ 
of various $\overline B\to PP$ modes.}
\begin{tabular}{|l|ccc|ccc|}
\hline
 Mode
      &${\cal B}^{\rm Exp}(10^{-6})$
      &${\cal B}^{\rm Fac}(10^{-6})$
      &${\cal B}^{\rm FSI}(10^{-6})$
      &${\cal A}^{\rm Exp}(\%)$
      &${\cal A}^{\rm Fac}(\%)$
      &${\cal A}^{\rm FSI}(\%)$
      \\
\hline
 $\ov B{}^0\to K^-\pi^+$
        & $19.4\pm0.6$
        & (16.0)
        & $20.1_{-0.3}^{+1.7}{}_{-2.5}^{+2.5}$
        & $-9.8^{+1.2}_{-1.1}$
        & $(-11.8)$
        & $-9.0_{-0.6}^{+2.0}{}_{-2.2}^{+2.0}$
        \\
 $\ov B {}^0\to \ov K {}^0\pi^0$
        & $9.8\pm0.6$
        & (7.2)
        & $9.2_{-0.2}^{+0.7}{}_{-1.2}^{+1.2}$
        & $-1\pm13^{*c}$ 
        & $(3.3)$
        & $-12.8_{-1.0}^{+2.2}{}_{-1.5}^{+1.7}$
        \\
 $\overline B {}^0\to \ov K {}^0\eta$
        & $1.0\pm0.3$
        & (0.9)
        & $1.4_{-0.1}^{+0.4}{}_{-0.4}^{+0.5}$
        & --
        & (10.7)
        & $-28.7_{-1.9}^{+8.0}{}_{-1.9}^{+3.3}$
        \\
 $\overline B {}^0\to \ov K {}^0\eta'$
        & $64.9\pm3.1$
        & (66.4)
        & $65.9_{-10.6}^{+6.9}{}_{-8.1}^{+9.2}$
        & $4.8\pm5.1$
        & ($0.2$)
        & $1.7_{-0.2}^{+0.8}{}_{-0.4}^{+0.3}$
        \\
        \hline
 $B^-\to \ov K{}^0\pi^-$
        & $23.1\pm1.0$
        & (18.0)
        & $22.5_{-1.1}^{+2.6}{}_{-0.7}^{+3.0}$
        & $0.9\pm2.5$
        & (0.3)
        & $-0.3_{-0.6}^{+0.7}{}_{-1.1}^{+1.2}$
        \\
 $B^-\to K^-\pi^0$
        & $ 12.9\pm 0.6 $
        & (10.1)
        & $12.4_{-0.2}^{+1.5}{}_{-1.6}^{+1.6}$
        & $5.0\pm 2.5 $
        & $(-11.8)$
        & $4.8_{-1.2}^{+1.4}{}_{-2.0}^{+1.9}$
        \\
 $B^-\to K^-\eta$
        & $2.7\pm 0.3$
        & $(1.4)$
        & $2.1_{-0.1}^{+0.6}{}_{-0.5}^{+0.6}$
        & $-27\pm 9$
        & (39.8)
        & $-27.3_{-3.0}^{+8.6}{}_{-6.3}^{+10.8}$
        \\
 $B^-\to K^-\eta'$
        & $70.2\pm 2.5$
        & (70.1)
        & $70.8_{-12.3}^{+6.6}{}_{-9.2}^{+10.3}$
        & $1.6\pm 1.9 $
        & $(-2.6)$
        & $-3.3_{-0.5}^{+1.0}{}_{-0.5}^{+0.5}$
        \\
        \hline
 $B^-\to \pi^-\pi^0$
        & $5.59^{+0.41}_{-0.40}$
        & $(5.18)$
        & $5.18_{-0.38}^{+0.55}{}_{-0.00}^{+0.00}$
        & $6\pm5$
        & $(-0.06)$
        & $-0.06_{-0.01}^{+0.00}{}_{-0.00}^{+0.00}$
        \\
 $B^-\to K^0 K^-$
        & $1.36^{+0.29}_{-0.27}$
        & $(1.22)$
        & $1.46_{-0.04}^{+0.35}{}_{-0.13}^{+0.15}$
        & $12^{+17}_{-18}$
        & $(-3.5)$
        & $12.8_{-12.8}^{+9.1}{}_{-17.8}^{+16.0}$
        \\
 $B^-\to \pi^-\eta$
        & $4.4\pm0.4$
        & $(4.10)$
        & $4.23_{-0.23}^{+0.59}{}_{-0.37}^{+0.34}$
        & $-16\pm7$
        & (19.7)
        & $-12.3_{-2.9}^{+4.1}{}_{-3.2}^{+3.5}$
        \\
 $B^-\to \pi^-\eta'$
        & $2.7^{+0.6}_{-0.5}$
        & (3.09)
        & $3.31_{-0.51}^{+0.19}{}_{-0.54}^{+0.65}$
        & $21\pm15$
        & (22.8)
        & $54.8_{-10.6}^{+5.3}{}_{-3.0}^{+1.7}$
        \\
\hline
 $\ov B {}^0\to \pi^+\pi^-$
        & $5.16\pm 0.22$
        & (6.65)
        & $5.30_{-0.49}^{+1.92}{}_{-0.40}^{+0.39}$
        & $38\pm15^{*d}$ 
        & (22.3)
        & $15.5_{-4.3}^{+10.2}{}_{-4.5}^{+4.6}$
        \\
 $\ov B {}^0\to \pi^0 \pi^0$
        & $1.55\pm0.35^{*a}$
        & (0.50)
        & $1.04_{-0.55}^{+0.12}{}_{-0.08}^{+0.10}$
        & $43^{+25}_{-24}$
        & $(-51.5)$
        & $48.3_{-33.1}^{+11.5}{}_{-13.1}^{+11.8}$
        \\
 $\ov B {}^0\to \eta\eta$
        & $0.8\pm0.4(<1.4)$
        & (0.21)
        & $0.46_{-0.11}^{+0.24}{}_{-0.08}^{+0.10}$
        & --
        & $(-11.7)$
        & $-50.7_{-12.4}^{+15.0}{}_{-16.3}^{+15.7}$
        \\
 $\ov B {}^0\to \eta \eta'$
        & $0.5\pm0.4(<1.2)$
        & (0.22)
        & $0.88_{-0.40}^{+0.39}{}_{-0.21}^{+0.24}$
        & --
        & $(-28.5)$
        & $-5.7_{-22.2}^{+9.5}{}_{-7.4}^{+7.8}$
        \\
 $\ov B {}^0\to \eta'\eta'$
        & $0.9_{-0.7}^{+0.8}(<2.1)$
        & (0.16)
        & $1.06_{-0.31}^{+1.16}{}_{-0.28}^{+0.36}$
        & --
        & (3.6)
        & $29.7_{-1.7}^{+26.2}{}_{-6.6}^{+8.3}$
        \\
 $\ov B {}^0\to K^+ K^-$
        & $0.15^{+0.11}_{-0.10}$
        & (0.09)
        & $0.10_{-0.02}^{+0.35}{}_{-0.06}^{+0.10}$
        & --
        & (0)
        & $71.0_{-41.4}^{+10.9}{}_{-15.6}^{+20.6}$
        \\
 $\ov B {}^0\to K^0\ov K^0$
        & $0.96^{+0.21}_{-0.19}$
        & (1.47)
        & $1.10_{-0.12}^{+0.46}{}_{-0.11}^{+0.12}$
        & $-58^{+73}_{-66}$
        & $(-9.0)$
        & $-37.8_{-37.1}^{+\,\,\,8.4}{}_{-15.0}^{+15.2}$
        \\
 $\ov B {}^0\to \pi^0 \eta$
        & $0.9\pm0.4(<1.5)$
        & $(0.26)$
        & $0.31_{-0.01}^{+0.05}{}_{-0.06}^{+0.06}$
        & --
        & (19.7)
        & $7.2_{-13.8}^{+11.5}{}_{-0.5}^{+0.4}$
        \\
 $\ov B {}^0\to \pi^0\eta'$
        & $1.2\pm0.7^{*b}$
        & (0.32)
        & $0.42_{-0.15}^{+0.02}{}_{-0.11}^{+0.13}$
        & --
        & (13.2)
        & $22.7_{-20.5}^{+\,\,\,7.7}{}_{-1.0}^{+1.0}$
        \\
        \hline
 $\ov B_s{}^0\to K^-\pi^+$
        & $5.00\pm1.25$
        & (4.72)
        & $4.81_{-0.39}^{+1.57}{}_{-0.22}^{+0.20}$
        & $39\pm17$
        & (33.4)
        & $26.6_{-5.2}^{+2.7}{}_{-4.7}^{+4.8}$
        \\
 $\ov B_s {}^0\to \ov K {}^0\pi^0$
        & --
        & (0.68)
        & $1.13_{-0.33}^{+0.24}{}_{-0.04}^{+0.05}$
        & --
        & $(-49.1)$
        & $45.5_{-12.6}^{+30.7}{}_{-10.5}^{+10.1}$
        \\
 $\ov B_s {}^0\to \ov K {}^0\eta$
        & --
        & $(0.28)$
        & $0.59_{-0.16}^{+0.10}{}_{-0.04}^{+0.04}$
        & --
        & $(2.0)$
        & $76.4_{-5.1}^{+14.9}{}_{-7.7}^{+6.0}$
        \\
 $\ov B_s {}^0\to \ov K {}^0\eta'$
        & --
        & $(2.33)$
        & $2.44_{-0.44}^{+0.14}{}_{-0.36}^{+0.42}$
        & --
        & $(2.5)$
        & $-14.6_{-21.8}^{+4.3}{}_{-4.2}^{+5.7}$
        \\
        \hline
 $\ov B_s {}^0\to \pi^+\pi^-$
        & $0.53\pm 0.51$
        & $(0.30)$
        & $0.86_{-0.19}^{+1.72}{}_{-0.85}^{+2.93}$
        & --
        & $(0)$
        & $-6.1_{-1.2}^{+9.7}{}_{-21.5}^{+56.4}$
        \\
 $\ov B_s {}^0\to \pi^0 \pi^0$
        & --
        & $(0.15)$
        & $0.43_{-0.10}^{+0.86}{}_{-0.43}^{+1.47}$
       & --
        & $(0)$
        & $-6.1_{-1.2}^{+9.7}{}_{-21.5}^{+56.4}$
        \\
 $\ov B_s {}^0\to \eta\eta$
        & --
        & $(17.5)$
        & $20.2_{-1.2}^{+7.6}{}_{-4.5}^{+5.9}$
        & --
        & $(1.6)$
        & $-3.6_{-1.6}^{+2.6}{}_{-1.4}^{+1.9}$
        \\
 $\ov B_s {}^0\to \eta \eta'$
        & --
        & $(70.8)$
        & $63.6_{-9.2}^{+47.1}{}_{-9.7}^{+13.7}$
        & --
        & $(0.4)$
        & $0.2_{-0.1}^{+1.7}{}_{-1.0}^{+1.1}$
        \\
 $\ov B_s {}^0\to \eta'\eta'$
        & --
        & $(81.9)$
        & $99.1_{-72.3}^{+6.9}{}_{-13.4}^{+15.2}$
        & --
        & $(0.2)$
        & $0.0_{-3.5}^{+0.2}{}_{-0.3}^{+0.4}$
        \\
 $\ov B_s {}^0\to K^+ K^-$
        & $24.4\pm4.8$
        & $(24.7)$
        & $20.7_{-2.1}^{+11.5}{}_{-3.0}^{+3.3}$
        & --
        & $(-11.9)$
        & $-11.0_{-1.3}^{+3.1}{}_{-2.9}^{+2.7}$
        \\
 $\ov B_s {}^0\to K^0\ov K^0$
        & --
        & (25.4)
        & $20.4_{-1.8}^{+12.1}{}_{-3.4}^{+3.8}$
        & --
        & (0.3)
        & $2.2_{-0.3}^{+1.8}{}_{-1.1}^{+1.2}$
        \\
 $\ov B_s {}^0\to \pi^0 \eta$
        & --
        & (0.06)
        & $0.09_{-0.00}^{+0.03}{}_{-0.00}^{+0.00}$
        & --
        & $(3.9)$
        & $82.8_{-20.0}^{+5.5}{}_{-4.9}^{+4.2}$
        \\
 $\ov B_s {}^0\to \pi^0\eta'$
        & --
        & (0.09)
        & $0.13_{-0.00}^{+0.03}{}_{-0.01}^{+0.01}$
        & --
        & (37.5)
        & $93.9_{-15.5}^{+2.7}{}_{-4.4}^{+3.2}$
        \\
        \hline
 \end{tabular}
 \footnotetext[0]{\centerline{\hspace{-0.65cm}${}^*$$S$ factors of ${}^a$1.8, ${}^b$1.7 ${}^c$1.4 and ${}^d$2.4 are included in the
 uncertainties, respectively.}}
 \end{table}

In Table~\ref{tab:table-br}, we show the CP-average rates and
direct CP violations of $\ov B{}^0,B^-,\ov B_s\to PP$
decays~\cite{Chua:2007cm}. In the table, Fac
and FSI denote factorization 
and FSI results, respectively.

\begin{figure}[t!]
\centerline{\DESepsf(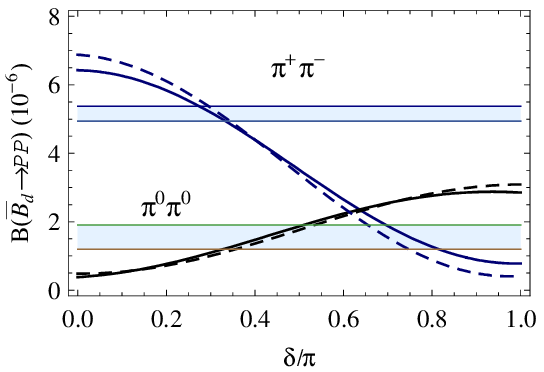 width
7cm)\hspace{1cm}\DESepsf(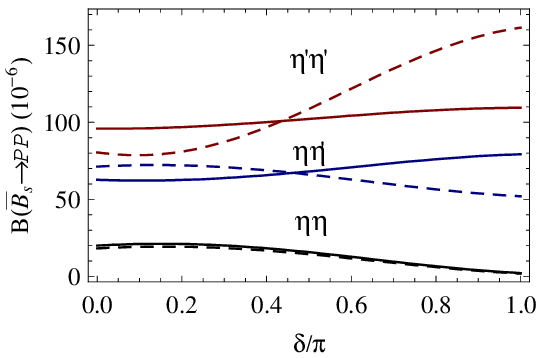 width 7cm)}
\centerline{\hspace{1cm}(a)\hspace{7.5cm}(b)}
\centerline{\DESepsf(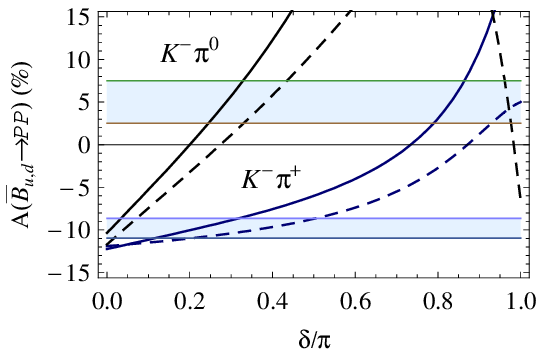 width 7.2cm)\hspace{1cm}\DESepsf(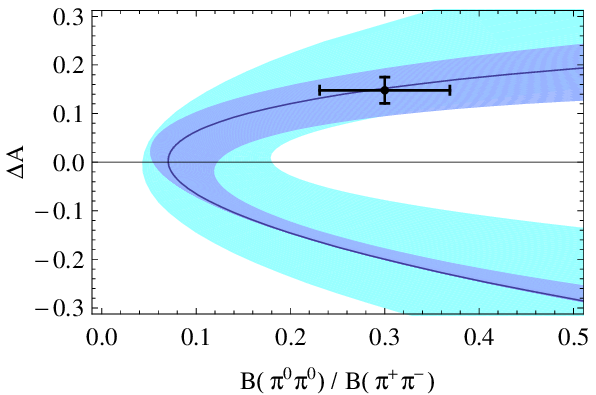 width 7.2cm)}
%
\centerline{\hspace{1cm}(c)\hspace{7.5cm}(d)}
\caption{(a) $\ov B {}^0\to \pi^+\pi^-$, $\pi^0\pi^0$ rates, (b)
$\ov B {}^0_s\to \eta^{(\prime)}\eta^{(\prime)}$ 
rates and (c) direct CP violations of $\ov B {}^0\to
K^-\pi^+$ and $B^-\to
K^-\pi^0$ 
versus the FSI phase $\delta$ are plotted. The solid (dashed) line
corresponds to the SU(3) (exchange-type U(3)) case. Bands are
1-$\sigma$ ranges of data. Theoretical uncertainties are not
shown. Note that the fitted $\delta/\pi$ is around $0.3$.
(d)~Correlation of the ratio $\B(\ov B{}^0\to\pi^0\pi^0)/\B(\ov
B{}^0\to\pi^+\pi^-)$ with the difference
$\Delta\A\equiv\A(K^-\pi^+)-\A(K^-\pi^0)$ is also plotted. The
light shaded area corresponds to the restricted SU(3) case, the
dark shaded area and the solid line correspond to the
exchange-type U(3) case.}
\label{fig:B0Br}\label{fig:Kpiacp}\label{fig:correlation}\label{fig:Bs}
\end{figure}

We discuss the rates in $\overline B {}^0$ and $B^-$ decays first.
As shown in Table~\ref{tab:table-br}, the residual FSI results
agree with data. After the residual FSI is turned on, some rates
are enhanced remarkably.  We see that $\ov B{}^0$ decays in the
$\Delta S=0$ transitions receive large contributions from the
residual FSI. In particular, through the residual FSI, $\ov
B{}^0\to\pi^+\pi^-$ and
$\pi^0\pi^0$ rates 
are reduced and enhanced roughly by factor $2$, respectively,
leading to a better agreement with data.
In Fig.~\ref{fig:B0Br}(a), we show the $\ov B{}^0\to\pi^+\pi^-$
and $\pi^0\pi^0$ rates versus the FSI phase $\delta$. We see that
$\ov B{}^0\to\pi^+\pi^-$ and $\pi^0\pi^0$ rates are reduced and
enhanced, respectively, as $\delta$ is increasing. Both rates
reach the measured ones at $\delta\sim0.3\pi$.

It is known that in order to have the $\pi^0\pi^0$ rate as large
as observed, we need a sizable color-suppressed tree
amplitude~\cite{largeC}. In the residual FSI, a large
color-suppressed tree contribution can be generated from the
exchange rescattering. As shown in the upper part of
Fig.~\ref{fig:re}, the color-allowed tree amplitude of the $\ov
B{}^0\to\pi^+\pi^-$ decay, the main FSI source in this sector, can
produce a color-suppressed tree amplitude for the $\ov
B{}^0\to\pi^0\pi^0$ decay through the exchange rescattering. At
the same time, the $\pi^+\pi^-$ rate is reduced as it rescatters.

\begin{table}[b!]
\caption{ \label{tab:table-S} Results on the time-dependent CP
asymmetry $S$ of various $\overline B_{d,s}\to PP$ modes.
}
\begin{tabular}{|l|ccc|ccc|}
\hline
 Mode
      &$S^{\rm Exp}(B_d)$ 
      &$S^{\rm Fac}(B_d)$ 
      &$S^{\rm FSI}(B_d)$ 
      &$S^{\rm Exp}(B_s)$ 
      &$S^{\rm Fac}(B_s)$ 
      &$S^{\rm FSI}(B_s)$ 
      \\
\hline
 $\ov B {}^0_{d,s}\to K_S\pi^0$
        & $0.58\pm0.17$
        & $(0.780)$
        & $0.778_{-0.037}^{+0.003}{}_{-0.013}^{+0.014}{}_{-0.002}^{+0.003}$
        & --
        & $(-0.315)$
        & $-0.155_{-0.147}^{+0.116}{}_{-0.047}^{+0.061}{}_{-0.164}^{+0.101}$
        \\
 $\ov B{}^0_{d,s}\to K_S\eta$
        & --
        & $(0.831)$
        & $0.769_{-0.050}^{+0.013}{}_{-0.039}^{+0.043}{}_{-0.001}^{+0.000}$
        & --
        & $(-0.137)$
        & $0.076_{-0.416}^{+0.255}{}_{-0.050}^{+0.031}{}_{-0.157}^{+0.091}$
        \\
 $\ov B{}^0_{d,s}\to K_S\eta'$
        & $0.60\pm0.07$
        & $(0.691)$
        & $0.682_{-0.002}^{+0.008}{}_{-0.004}^{+0.004}{}_{-0.000}^{+0.000}$
        & --
        & $(-0.174)$
        & $0.001_{-0.109}^{+0.046}{}_{-0.0848}^{+0.077}{}_{-0.001}^{+0.001}$
        \\
        \hline
 $\ov B{}^0_{d,s}\to \pi^+\pi^-$
        & $-0.65\pm 0.07$
        & $(-0.591)$
        & $-0.542_{-0.005}^{+0.088}{}_{-0.034}^{+0.038}{}_{-0.074}^{+0.139}$
        & --
        & $(0.143)$
        & $0.095_{-0.014}^{+0.055}{}_{-0.942}^{+0.109}{}_{-0.001}^{+0.002}$
        \\
 $\ov B{}^0_{d,s}\to \pi^0 \pi^0$
        & --
        & $(0.854)$
        & $0.484_{-0.114}^{+0.425}{}_{-0.109}^{+0.096}{}_{-0.096}^{+0.145}$
        & --
        & $(0.143)$
        & $0.095_{-0.014}^{+0.055}{}_{-0.942}^{+0.109}{}_{-0.001}^{+0.002}$
        \\
 $\ov B{}^0_{d,s}\to \eta\eta$
        & --
        & $(-0.985)$
        & $-0.308_{-0.237}^{+0.122}{}_{-0.110}^{+0.144}{}_{-0.089}^{+0.160}$
        & --
        & $(-0.041)$
        & $-0.057_{-0.002}^{+0.029}{}_{-0.017}^{+0.016}{}_{-0.004}^{+0.003}$
        \\
 $\ov B{}^0_{d,s}\to \eta \eta'$
        & --
        & $(-0.945)$
        & $-0.946_{-0.036}^{+0.015}{}_{-0.016}^{+0.020}{}_{-0.016}^{+0.034}$
        & --
        & $(-0.006)$
        & $-0.016_{-0.007}^{+0.016}{}_{-0.003}^{+0.005}{}_{-0.002}^{+0.001}$
        \\
 $\ov B{}^0_{d,s}\to \eta'\eta'$
        & --
        & $(-0.901)$
        & $-0.917_{-0.024}^{+0.089}{}_{-0.021}^{+0.030}{}_{-0.000}^{+0.001}$
        & --
        & $(0.031)$
        & $0.048_{-0.014}^{+0.013}{}_{-0.003}^{+0.003}{}_{-0.000}^{+0.000}$
        \\
 $\ov B{}^0_{d,s}\to K^+ K^-$
        & --
        & $(-0.920)$
        & $-0.630_{-0.289}^{+0.091}{}_{-0.187}^{+0.521}{}_{-0.046}^{+0.085}$
        & --
        & $(0.194)$
        & $0.195_{-0.035}^{+0.019}{}_{-0.021}^{+0.017}{}_{-0.004}^{+0.005}$
        \\
 $\ov B{}^0_{d,s}\to K^0\ov K^0$
        & $-0.38^{+0.69}_{-0.77}\pm0.09$
        & $(-0.110)$
        & $0.327_{-0.283}^{+0.264}{}_{-0.068}^{+0.072}{}_{-0.011}^{+0.002}$
        & --
        & $(0.005)$
        & $-0.010_{-0.010}^{+0.023}{}_{-0.005}^{+0.007}{}_{-0.002}^{+0.001}$
        \\
        & $-1.28^{+0.80}_{-0.73}{}^{+0.11}_{-0.16}$
        &
        &
        &
        &
        &
        \\
 $\ov B{}^0_{d,s}\to \pi^0 \eta$
        & --
        & $(0.019)$
        & $0.057_{-0.145}^{+0.151}{}_{-0.012}^{+0.011}{}_{-0.004}^{+0.000}$
        & --
        & $(0.691)$
        & $0.140_{-0.230}^{+0.175}{}_{-0.007}^{+0.008}{}_{-0.025}^{+0.044}$
        \\
 $\ov B{}^0_{d,s}\to \pi^0\eta'$
        & --
        & $(0.043)$
        & $0.084_{-0.124}^{+0.064}{}_{-0.018}^{+0.016}{}_{-0.003}^{+0.0001}$
        & --
        & $(0.816)$
        & $0.135_{-0.145}^{+0.169}{}_{-0.096}^{+0.095}{}_{-0.037}^{+0.065}$
        \\
         \hline
\end{tabular}
\end{table}

We turn to results for direct CP asymmetries in $\ov B{}^0,B^-\to
PP$ decays. In general, the residual FSI has a large impact on
direct CP violations of many modes. We concentrate on the modes
that lead to the $K\pi$ puzzle. From Table~\ref{tab:table-acp} we
see that before the residual FSI is turned on (i.e. taking
$\Sc_{res}=1$), we have $\A(\ov B{}^0\to
K^-\pi^+)\simeq\A(B^-\to K^-\pi^0)\simeq-0.12$. 
After turning on the residual FSI $(\Sc_{res}\neq 1)$, the
asymmetry $\A(\ov B{}^0\to K^-\pi^+)$ changes from $\sim -0.12$ to
$\sim -0.09$, while $\A(B^-\to K^-\pi^0)$ changes from $\sim
-0.12$ to $\sim +0.05$, reproducing the experimental results. The
residual FSI has a more prominent effect on $\A(B^-\to K^-\pi^0)$,
and, hence, it is capable of lifting the degeneracy of $\A(B^-\to
K^-\pi^0)$ and $\A(\ov B{}^0\to K^-\pi^+)$.
It is known that a sizable and complex color-suppressed tree
amplitude in the $B^-\to K^-\pi^0$ decay can solve the $K\pi$
puzzle~\cite{largeC}. As depicted in Fig.~\ref{fig:re}, a
color-suppressed tree amplitude in the $K^-\pi^0$ mode can be
generated from the exchange rescattering of $B^-\to
K^-\eta^{(\prime)}$ color-allowed tree amplitudes. 
The rescattering leads to the desired large and complex
color-suppressed amplitude in the $K^-\pi^0$ mode and resolves the
$K\pi$ direct CP violation puzzle without the need of introducing
any new physics contribution.

As noted before, the exchange rescattering is also responsible for
the enhancement of the $\ov B{}^0\to \pi^0\pi^0$ rate. In
Fig.~\ref{fig:correlation}(d), we show a two-dimensional plot,
exhibiting the correlation of the ratio $\B(\ov
B{}^0\to\pi^0\pi^0)/\B(\ov B{}^0\to\pi^+\pi^-)$ with the
difference $\Delta\A\equiv\A(\ov B{}^0\to K^-\pi^+)-\A(B^-\to
K^-\pi^0)$. The light shaded area corresponds to the restricted
SU(3) case, the dark shaded area and the solid line correspond to
the exchange-type U(3) case. We clearly see that the data can be
easily reproduced and the exchange rescattering is responsible for
generating sizable and complex color-suppressed tree amplitudes
that account for the difference $\Delta\A$ and the $\B(\ov
B{}^0\to\pi^0\pi^0)/\B(\ov B{}^0\to\pi^+\pi^-)$ ratio at the same
time.

We also note that the direct CP violation of $B^-\to \pi^-\pi^0$
is very small and does not receive any contribution from the
residual rescattering, since it can only rescatter into itself.
The $\A(B^-\to\pi^-\pi^0)$ measurement remains as a clean way to
search for new physics effects~\cite{Cheng:2004ru}.

We now turn to $B_s$ decays. 
Comparing to data, we see from Table~\ref{tab:table-Bs} that
the $\B(\ov B_s\to K^-\pi^+)$, $\B(\ov B_s\to K^+K^-)$ rates agree
well with data, while the data on $\B(\ov B_s\to \pi^+\pi^-)$ and
$\A(\ov B_s\to K^-\pi^+)$ can be reproduced, the results have
large uncertainties.
We expect the residual FSI to have sizable contributions to
various $\ov B_s\to PP$ decay rates (see the lower diagram in
Fig.~\ref{fig:re}). The plot of $\ov B_s\to
\eta^{(\prime)}\eta^{(\prime)}$ rate versus $\delta$ is shown in
Fig.~\ref{fig:Bs}(b). The $\ov B{}^0_s\to\eta'\eta'$ rate is
enhanced and reaches $1.0\times 10^{-4}$, which can be checked
soon. The $\ov B_s\to \ov K{}^0\pi^0,\ov K{}^0\eta$ modes are also
quite sensitive to the residual rescattering. Similar to $\ov
B_{u,d}$ cases, the residual FSI also has large impacts on many
$\A(\ov B_s\to PP)$.

Results on time-dependent CP-asymmetries $S$ are given in
Table~\ref{tab:table-S}~\cite{Chua:2007cm}. The last uncertainty
comes from the variation of
$\gamma/\phi_3=(67.6^{+2.8}_{-4.5})^\circ$~\cite{CKMfitter}.
We fit to data on mixing induced CP asymmetries.
For $\ov B{}^0$ decays, we define $\Delta S\equiv \sin2\beta_{\rm
eff}-\sin2\beta_{c\bar c K}$, where $\sin2\beta_{\rm eff}=-\eta_f
S(f)$ with $\eta_f$ the CP eigenvalue of the state $f$. Comparing
with the recent value of $\sin2\beta_{c\bar c
K}=0.671\pm0.024$~\cite{data} as measured in $B^0\to K$ +
charmonium modes, we obtain:
 \be
 \Delta S(K_S\pi^0)=0.107^{+0.028}_{-0.046},\quad
 \Delta S(K_S\eta)=0.098^{+0.051}_{-0.068},\quad
 \Delta S(K_S\eta')=0.011^{+0.026}_{-0.024}.
 \en
The $\Delta S(K_S\eta')$, is a promising test of the SM. These
$\Delta S(B_d)$ agrees with those found in
\cite{Cheng:2004ru,Beneke:2005pu,review,CharmingP}.

For $\ov B{}^0_s$ decays, the $S$ contributed from $\ov
B{}^0_s$--$B_s^0$ mixing itself is around $-0.036$. Hence, for
penguin dominated $b\to s$ transition, we do not expect the
corresponding $|S|$ to be much larger than ${\cal O}(0.05)$.
Indeed, the predicted $|S|$ as shown in Table~\ref{tab:table-S}
for $\ov B {}^0_s\to \eta\eta$, $\eta'\eta'$ and $\eta\eta'$
decays are all below $0.06$. Given the recent interesting
preliminary results in the $B_s$ phase~\cite{UTfit,data}, it will
be very useful to search for $S$ in these $B_s$ charmless decays.

\begin{acknowledgments}
%
Work supported in part by the National Science Council of R.O.C.
under Grant No. NSC 97-2112-M-033-002-MY3.
\end{acknowledgments}

\end{document}